\documentclass[twocolumn,amsmath,aps,superscriptaddress,showpacs,floatfix,a4paper]{revtex4}

\usepackage{graphicx}
\usepackage{placeins}

\begin{document}
\title{The interplay between hydrodynamic and Brownian fluctuations in sedimenting
colloidal suspensions}
\author{J.\ T.\ Padding}
\affiliation{Computational Biophysics, University of Twente, PO Box 217, 7500 AE, Enschede, The Netherlands}
\affiliation{Department of Chemistry, Cambridge University,
           Lensfield Road, Cambridge CB2 1EW, United Kingdom}
\author{A.\ A.\ Louis}
\affiliation{Department of Chemistry, Cambridge University,
           Lensfield Road, Cambridge CB2 1EW, United Kingdom}
\affiliation{Rudolf Peierls Centre for Theoretical Physics,
           1 Keble Road, Oxford OX1 3NP, United Kingdom}
\date{\today}

\begin{abstract}
We apply a hybrid Molecular Dynamics and mesoscopic simulation technique to
study the steady-state sedimentation of hard sphere particles for Peclet numbers (Pe) ranging
from $0.08$ to $12$. Hydrodynamic back-flow causes a reduction of the average
sedimentation velocity relative to the Stokes velocity. We find that this
effect is independent of Pe number. Velocity fluctuations show the expected
effects of thermal fluctuations at short correlation times. At longer times,
non-equilibrium hydrodynamic fluctuations are visible, and their
character appears to be independent of the thermal fluctuations. The
hydrodynamic fluctuations dominate the diffusive behavior even for
modest Pe number, while conversely the short-time fluctuations are
dominated by thermal effects for a surprisingly large Pe numbers.
Inspired by recent experiments, we also study finite sedimentation in a
horizontal planar slit. In our simulations distinct lateral patterns emerge,
in agreement with observations in the experiments.
\end{abstract}


\maketitle

\section{Introduction}

The steady-state sedimentation of spheres in a viscous medium at low
Reynolds (Re) number is an important model problem in non-equilibrium
statistical mechanics, exhibiting subtle and interesting physics
\cite{Dhont,Ramaswamy,Russ89}.  Some properties are relatively
straightforward to determine. For example, the sedimentation velocity
$V_S^0$ of a single sphere was first  calculated over 150 years ago by
George Gabriel Stokes\cite{Stokes} to be $V_S^0 = \frac{2}{9} g
a^2\left(\rho_c - \rho\right)/\eta$,
where $a$ and $\rho_c$ are the radius the density of the sphere, $g$
is the gravitational acceleration, and $\rho$ and $\eta$ are the
density and viscosity of the fluid.  On the other hand, even the first
order effect of finite volume fraction $\phi =
\frac{4}{3} \pi n_c a^3$ ($n_c$ is the particle number density), was
not calculated until 1972 when Batchelor \cite{Batchelor} showed that
\begin{equation}\label{eq1.2}
V_S = V_S^0 \left(1 - 6.55 \phi + ({\cal O}\phi^2) \right).
\end{equation}
The effect of a finite volume fraction on sedimentation is dominated
by long-ranged hydrodynamic forces that decay with interparticle
distance $r$ as slowly as $r^{-1}$. These forces are hard to treat
analytically because they can easily lead to spurious divergences.
Eq.~(\ref{eq1.2}) also highlights the strong effect of the
hydrodynamic forces. For example, a naive application of this lowest
order result would suggest that all sedimentation should stop at $\phi
\approx 0.15$. Of course this is not true since there are important
higher order corrections in $\phi$ whose calculation remains an active
topic of research~\cite{Haya95}.

If the influence of hydrodynamics on the average sedimentation
velocity at finite volume fraction is non-trivial to calculate, then
the fluctuations around that average would appear even more formidable
to determine.  In a remarkable paper, Caflish and Luke \cite{CaflishLuke} used a simple
scaling argument to predict that,
for a 
homogeneous suspension, the velocity fluctuations
$\delta V = V-V_S$ should diverge as $\left\langle\delta V^2 \right\rangle \sim L$,
where $L$ is the smallest container
size. This surprising result stimulated much theoretical and
experimental work, as well as no small amount of
controversy \cite{Ramaswamy}. Particle velocimetry experiments
clearly show the existence of large-scale velocity fluctuations, which
manifest as ``swirls'' \cite{Nico95,Segr97,Segr01,Bern02,Tee02}.
The experiments of Nicolai \textit{et al.} \cite{Nico95} and Segr\'e \textit{et al.} \cite{Segr97} suggest
that while for small containers the velocity fluctuations do indeed
grow linearly in $L$, for larger containers the velocity fluctuations
saturate (see however~\cite{Bern02,Bren99}).  The reasons (if any) that this
should be observed have been the subject of sustained theoretical
debate. It was shown by Koch and Shaqfeh \cite{Koch91} that
hydrodynamic interactions can be screened if the colloids exhibit
certain long-ranged correlations reminiscent of those found for
electrostatic systems. A number of theories have been proposed to generate
such correlations in the bulk, including a coupled
convective-diffusion model by Levine \textit{et al.} \cite{Levi98} that generates a
noise-induced phase-transition to a screened phase.  Another class of
theories focuses on the container walls. For example Hinch \cite{Hinc88} has
argued that the bottom of a vessel will act as a sink
for fluctuations, a prediction that appears to be confirmed by
computer simulations \cite{Ladd02,Nguy05}. Other authors have
emphasized the importance of
stratification \cite{Segr02,Tee02,Much03,Much04,Segr04} and polydispersity \cite{Berg03,Nguy05}.

Most of the theoretical studies of sedimentation described above have
focused on the non-Brownian limit where thermal fluctuations are
negligible.  This can be quantified by defining the Peclet number (Pe)
\begin{equation}
\mathrm{Pe} = \frac{V_S^0 a}{D_0} = \frac{M_b g a}{k_B T},
\label{defPe}
\end{equation}
where $D_0$ is the equilibrium self-diffusion constant and
$M_b = \frac43 \pi a^3(\rho_c-\rho)$ is the particle's buoyant
mass.  The non-Brownian limit then corresponds to ${\rm Pe}=\infty$.
Because Pe scales as $\left(\rho_c -\rho\right) a^4$, the very large Pe numbers needed
to approximate the non-Brownian limit are easily achieved by
increasing particle size.

The Pe number is directly related to the gravitational length $l_g =
k_B T/(M_b g) = a/\mathrm{Pe}$. For this reason, the criterion
$\mathrm{Pe}\leq 1$ is often used to define the colloidal regime
since, roughly speaking, one would expect from the barometric law that
particles would then be dispersed throughout the solution.  For
example, for polystyrene spheres in water $\mathrm{Pe} = {\cal O}(1)$
for $a \approx 1 \mu$m.  In experiments, the density difference
$\Delta \rho = \rho_c - \rho$ can be adjusted by density matching so
that the Pe number can also be tuned quite accurately for a given $a$.

In contrast to most previous theoretical and computational studies,
which have focused on the non-Brownian $\mathrm{Pe} =\infty$ limit, in this
paper we study steady state sedimentation at the moderate Pe numbers
relevant for the colloidal regime. In this regime the particles
experience both (random) thermal fluctuations and (deterministic)
hydrodynamic fluctuations, and a key question will be how these two
kinds of fluctuations interact.

We employ Stochastic Rotation Dynamics (SRD) \cite{Male99,IhleKroll,Padd06}
to describe the solvent, and a Molecular Dynamics scheme to propagate the
colloids. Such a hybrid technique was first
employed by Malevanets and Kapral \cite{Male00}, and recently used to
study colloidal sedimentation by ourselves \cite{PaddingLouis} and by Hecht \textit{et al.} \cite{Hech05}.
In section \ref{sec_method} we briefly recap the salient details of our simulation method.

In section \ref{sec_average} we study the average sedimentation velocity.
Our principle finding is that this follows exactly the same trend with volume fraction
$\phi$ as found for the $\mathrm{Pe}=\infty$ non-Brownian limit. In
other words, the effects of backflow are completely dominated by the hydrodynamic interactions
(HI), even when the Brownian forces are, on average, much stronger.   
In sections \ref{sec_space} and \ref{sec_time} we investigate in some detail the velocity fluctuations $\langle
\delta V^2 \rangle$. We find that the thermal and hydrodynamic
fluctuations appear to act independently of each other. Their
effects are additive, at least in the accessed simulation regime,
where the hydrodynamic fluctuations are unscreened.  Some of these
results have appeared earlier in a short letter \cite{PaddingLouis},
but here they are treated in much more detail.
In section \ref{sec_diffusion} we calculate the self-diffusion coefficient, highlighting the 
effects of hydrodynamic dispersion.
In section \ref{sec_finite} we briefly consider the case of finite sedimentation in a horizontal 
planar slit. We observe distinct lateral patterns, in agreement with recent laser scanning confocal microscopy.
In section \ref{sec_discussion} we discuss the importance of thermal fluctuations over hydrodynamic fluctuations.
Finally, in section \ref{sec_conclusion} we present our conclusions.

\section{Hybrid MD-SRD coarse-grained simulation method}
\label{sec_method}

 The time- and length-scale differences between colloidal
 and solvent particles are enormous: a typical colloid of diameter 1 $\mu$m will displace on the
 order of $10^{10}$ water molecules.
Clearly, some form of {\em coarse-graining} of the solvent is
necessary. In this paper we use SRD to efficiently describe the dynamics
of the solvent.  The colloids are coupled to the solvent
through explicit interaction
potentials. We have recently performed an extensive validation of
this method
\cite{Padd06}. We will therefore only reproduce the most important
conclusions.

\subsection{Solvent-solvent interactions}
 
In SRD, the solvent is represented by a large number $N_f$ of particles of
mass $m_f$. We will call these {\em fluid} particles,
with the caveat that, however tempting, they should not be viewed as some kind of
composite particles or clusters made up of the underlying molecular fluid.
The particles are merely a convenient computational device to facilitate the coarse-graining
of the fluid properties \cite{Padd06}.

In the first step, the positions and velocities of the fluid particles
are propagated by integrating Newton's equations of motion. The forces
on the fluid particles are generated by external forces generated by gravity, walls, or
colloids. Direct forces between pairs of fluid particles are, however,
excluded; this is the main reason for the efficiency of the
method. After propagating the fluid particles for a time $\Delta t_c$,
the second step of the algorithm simulates the collisions between
fluid particles. The system is partitioned into cubic cells of volume
$a_0^3$. The velocities relative to the center of mass velocity
$\mathbf{v}_{\mathrm{cm}}$ of each separate cell are then rotated:
\begin{equation}
\mathbf{v}_i \mapsto \mathbf{v}_{\mathrm{cm}} + \mathbf{R}\left(
\mathbf{v}_i - \mathbf{v}_{\mathrm{cm}} \right)\mbox{.}
\end{equation}
$\mathbf{R}$ is a rotation matrix which rotates velocities by a
fixed angle $\alpha$ around a randomly oriented axis.
The aim of the collision step is to transfer momentum between the fluid particles.
The rotation procedure can thus be viewed as a coarse-graining
of particle collisions over time and space. Because mass, momentum, and
energy are conserved locally, the correct (Navier-Stokes) hydrodynamic
equations are captured in the continuum limit, \textit{including} the effect
of thermal noise \cite{Male99}.


Ihle and Kroll \cite{IhleKroll} pointed out that at low temperatures or 
small collision times $\Delta t_c$ the transport coefficients of SRD 
show anomalies. These anomalies are caused by the fact that fluid particles
in a given cell can remain in that cell and participate in several collision
steps. They showed that under these circumstances the assumption of molecular
chaos and Galilean invariance are incorrect. They also showed how the
anomaly can be entirely cured by applying a random shift of the cell
coordinates before the collision step.
It is then possible to analytically
calculate the shear viscosity of the SRD fluid \cite{Yeomans}. 
Such expressions are very useful because they enable us to efficiently tune the viscosity
of the fluid, without the need of trial and error simulations.

\subsection{Colloid-colloid and colloid-solvent interactions}

In the simulation, colloidal spheres of mass $M$ are propagated through the Velocity
Verlet algorithm \cite{AllenTildesley} with a time step $\Delta t_{\mathrm{MD}}$.
The colloids are embedded in the fluid, and interact with the fluid particles through a 
repulsive (Weeks-Chandler-Andersen) potential:
\begin{equation}
\varphi_{cf}(r) = \left\{
\begin{array}{ll}
	4 \epsilon \left[ \left(\frac{\sigma_{cf}}{r}\right)^{12} - \left(\frac{\sigma_{cf}}{r}\right)^6 + \frac{1}{4} \right] & (r\leq 2^{1/6}\sigma_{cf}) \\
	0 & (r> 2^{1/6}\sigma_{cf})
\end{array}
\right.
\label{eq_phicf}
\end{equation}
The colloid-colloid interaction is represented by a similar, but steeper, repulsive potential:
\begin{equation}
\varphi_{cc}(r) = \left\{
\begin{array}{ll}
	4 \epsilon \left[ \left(\frac{\sigma_{cc}}{r}\right)^{48} - \left(\frac{\sigma_{cc}}{r}\right)^{24} + \frac{1}{4} \right] & (r\leq 2^{1/24}\sigma_{cc}) \\
	0 & (r> 2^{1/24}\sigma_{cc})
\end{array}
\right.
\end{equation}
The latter potential makes the colloids hard enough to approximate
hard spheres, yet smooth enough to enable accurate integration of the
equations of motion with a time step $\Delta t_{\mathrm{MD}}$ close to the collision time interval $\Delta t_c$.

Because the surface of a colloid is never perfectly smooth, collisions
with fluid particles transfer angular as well as linear
momentum. These interactions may be approximated by stick boundary
conditions.  We have studied several implementations of stick boundary
conditions for spherical colloids \cite{Padd05} and derived a version
of stochastic boundary conditions which reproduce linear and
angular momentum correlation functions that agree with Enskog theory
for short times and hydrodynamic mode-coupling theory for long
times. Nevertheless, in this paper we use the radial interactions
described in Eq.~(\ref{eq_phicf}). These do not transfer angular
momentum to a spherical colloid and so induce effective slip boundary
conditions. For many of the hydrodynamic effects we will discuss here
the difference with stick boundary conditions is quantitative, not
qualitative, and also well understood. For example, we have confirmed
that the flowfield around a single sedimenting sphere decays,
to first order, like $a/(2r)$ for a slip boundary sphere \cite{Padd06},
whereas it decays like $3a/(4r)$ for a stick boundary sphere.

To avoid (uncontrolled) depletion forces, we routinely choose the
colloid-fluid interaction range $\sigma_{cf}$ slightly below half the
colloid diameter $\sigma_{cc}/2$ \cite{Padd06}. 
There is no a-priori reason why the hydrodynamic radius should
be exactly half the particle-particle hard-core diameter for a
physical system. For charged systems, for example, 
the difference may be substantial. An additional advantage 
of this choice
is that more fluid particles will fit in the space between two colloids, and
consequently lubrication forces will be more accurately
represented. We have confirmed that with our parameters SRD resolves the
analytically known lubrication forces down to gap widths as small as $a/5$.
The agreement at small distances is caused also by repetitive collisions of the
fluid particles trapped between the two surfaces.
But at some point the lubrication force will break down.
An explicit correction could be applied to correctly resolve
these forces for very small distances, as was implemented by
Nguyen and Ladd \cite{Nguy02} for Lattice Boltzmann dynamics. However, in this paper our
choice of $\sigma_{cf}$ is small enough for SRD to sufficiently
resolve lubrication forces up to the point where the direct
colloid-colloid interactions start to dominate \cite{Padd06}.

\subsection{Time scales and hydrodynamic numbers}

Many different time scales govern the physics of a colloid of mass $M$ embedded in a solvent.
Hydrodynamic interactions propagate
by momentum diffusion and also by sound. The sonic time is the time it takes a sound wave
to travel the radius of a colloid, $t_{cs} = a/c_s$, where $c_s$ is the speed of sound.
The kinematic time, on the other hand, is the time it takes momentum to diffuse over the radius of a colloid,
$\tau_{\nu} = a^2/\nu$, where $\nu$ is the kinematic viscosity of the solvent.
For a colloid of radius $a = 1 \mu$m in water, $\tau_{cs} \approx 10^{-9}$ s and $\tau_{\nu} \approx 10^{-6}$ s.

The next time scale is the Brownian
time $\tau_B = M/\xi_S$, where $\xi_S = 6 \pi \eta a$ is the Stokes friction for stick boundary
conditions, or $4 \pi \eta a$ for slip boundary conditions. It
measures the time for a colloid to lose memory of its velocity (see, however, \cite{Padd06}).
The most relevant time scale for Brownian motion is the diffusion time $\tau_D = a^2/D_0$,
which measures how long it takes for a colloid to diffuse over a distance $a$ in the absence of flow.
For a colloid of $a=1 \mu$m in water, $\tau_D \approx 5$ s.

When studying sedimentation, the Stokes time is the time it takes a single
colloid to advect over its own radius, $t_S = a/V_S^0$. The Stokes time and the diffusion time are related
by the Peclet number: $\mathrm{Pe} = \tau_D / t_S$.
If $\mathrm{Pe} \gg 1$, then the colloid moves convectively over a
distance much larger than its radius $a$ in the time $\tau_D$ that it
diffused over the same distance. 
For $\mathrm{Pe} \ll 1$, on the
other hand, the opposite is the case, and the main transport mechanism
is diffusive. 
It is sometimes thought that for low Pe numbers
hydrodynamic effects can safely be ignored, but this is not always
true, as we will show.

In summary, in colloidal suspensions we encounter a range of time
scales, ordered like $t_{cs} < \tau_B <
\tau_{\nu} < (\tau_D, t_S)$, where $t_S$ may be smaller or larger than
$\tau_D$ depending on Pe, and where we have assumed $\rho_c \approx
\rho$ to justify $\tau_B < \tau_{\nu}$.
The entire range of time scales can span more than 10 orders of
magnitude.
Thankfully, it is not necessary to exactly reproduce each
of the different time scales in order to achieve a correct
coarse-graining of colloidal dynamics.  We can ``telescope down''
\cite{Padd06} the relevant time scales to a hierarchy which is
compacted to maximize simulation efficiency, but sufficiently
separated to correctly resolve the underlying physical behavior.
Keeping the relevant time-scales an separated by about an order of
magnitude should suffice. 

Similar arguments can be made for various hydrodynamic
numbers. For example, the Re number of sedimenting
colloidal particles is normally very low, on the order of $10^{-5}$ or
less.  But there is no need to take such a low value since many
relative deviations from the zero-Re Stokes regime scale with
Re$^2$.  Exactly how big an error one makes 
depends on what one is investigating, but for our
purposes we will take Re $\leq 0.4$ as an upper
bound. We have shown \cite{Padd06} that for the
friction on a sphere inertial effects are unimportant up to Re $\approx 1$.
We note that our upper bound on Re also ensures that the time
hierarchy condition $\tau_{\nu} < (t_S,
\tau_D)$ is fulfilled.  In principle Pe can be whatever we like as
long as Re remains low and the hierarchy is obeyed.

In our  simulations we choose an average number of fluid
particles per collision volume equal to $\gamma = 5$, a collision
interval $\Delta t_c = 0.1$ (in units of $t_0 = a_0 \sqrt{m_f/k_BT}$),
and a rotation angle $\alpha = \pi/2$, leading to a kinematic viscosity $\nu = 0.5\ a_0^2/t_0$.
We choose a colloidal mass $M = 125 m_f$, and interaction parameters $\sigma_{cf} = 2 a_0$, $\sigma_{cc} = 4.3
a_0$, and $\epsilon = 2.5k_BT$.  We have verified that this
choice leads to a small relative error in the full velocity
field, and that we can quantitatively calculate the the observed
friction on a colloid \cite{Padd06}. Note that this friction
is somewhat lower than expected on the basis of a hydrodynamic radius
set equal to $\sigma_{cf} = 2a_0$. This is due to additional Enskog
friction effects, where the different contributions to the friction
add ``in parallel'', as explained in Ref.~\cite{Padd06}.
The resulting effective hydrodynamic radius $a = 1.55 a_0$ will be used
throughout this paper. The time scales in our simulations are well separated: $t_{cs} = 1.2 t_0$,
$\tau_B = 2.5 t_0$, $\tau_{\nu} = 4.8 t_0$, and $\tau_D = 120 t_0$.

\section{Average sedimentation velocity}
\label{sec_average}

Sedimentation simulations were performed in a periodic box of
dimensions $L_x = L_y = 32a_0$ and $L_z = 96a_0$ (approx. $21 \times 21 \times 62 a$),
with $N = 8$ to 800 colloids.  The number of SRD particles was adjusted so that the free
volume outside the colloids contained an average of 5
particles per coarse-graining cell volume $a_0^3$. This corresponds to
a maximum of $N_f \approx 5 \times 10^5$ SRD particles.  A gravitational field $g$,
applied to the colloids in the $z$ direction, was varied to produce
different Peclet numbers, ranging from $\mathrm{Pe} = 0.08$ to
$\mathrm{Pe} = 12$. At the same time the Reynolds numbers ranged from
$\mathrm{Re} = 0.003$ to 0.4.  Note that the absence of a wall at the
top and bottom of the simulation box necessitates an additional
constraint to keep the center-of-mass of the entire system fixed.

Right after the simulations start, the colloidal positions and
velocities have not yet acquired their steady-state distributions. We
monitored block averages (in time) of the sedimentation velocity and
the behavior of sedimentation velocity fluctuations, which will be
discussed in the next section. We verified that there was no drift in
these properties after about $100$ Stokes times $t_S$, corresponding
to sedimentation down the height of about two periodic boxes. The
absence of any drift indicated that the suspensions were now in steady
state.

The simulations were subsequently run between 200 $t_S$ for $\mathrm{Pe}=0.08$ to 30,000 $t_S$ for $\mathrm{Pe}=12$.
To check that our system is large enough, we performed some runs for 1.5 and 2 times the box size described above,
finding no significant changes in our conclusions.

\begin{figure}
  \centerline{\scalebox{0.5}{\includegraphics{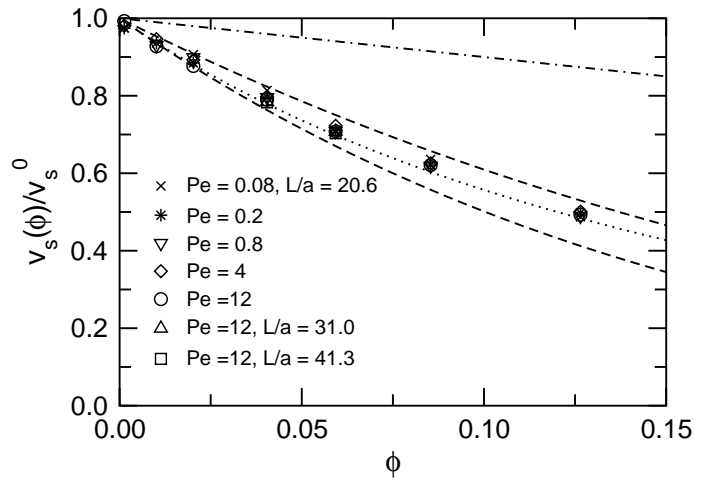}}}
\caption{Average sedimentation velocity, $V_S$ normalized by the Stokes
velocity $V_S^0$, as a function of volume fraction $\phi$ for various Peclet
numbers and system sizes. Dashed lines correspond to the
semi-empirical Richardson-Zaki law $(1-\phi)^n$, with $n = 4.7$ for the upper
and $n=6.55$ for the lower line. The dotted line is another
theoretical prediction taking higher order HI into
account\protect~\cite{Haya95}. Ignoring hydrodynamics leads to
$V_S/V_S^0=1-\phi$ (dash-dotted line).
\label{fig1}
}
\end{figure}
\begin{figure}
  \centerline{\scalebox{0.45}{\includegraphics{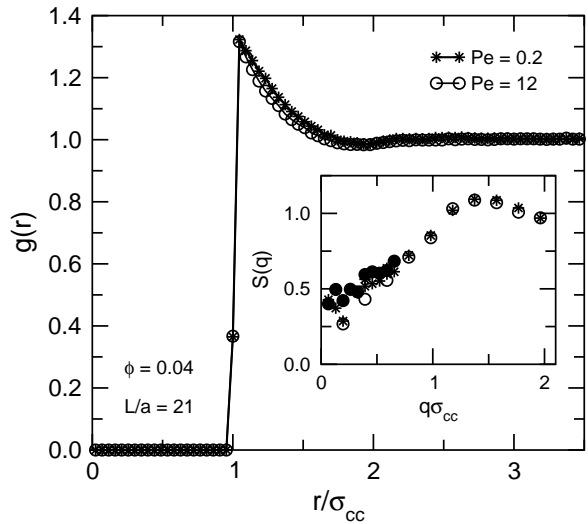}}}
\caption{Main plot: colloid radial distribution function $g(r)$ for $\phi = 0.04$
at low (0.2) and high (12) Peclet number.  There is no significant difference between the two $g(r)$'s.
Inset: structure factor for the same systems. At Pe = 12, small deviations are found for perpendicular (open
circles) and parallel (closed circles) wave vectors.
\label{fig2}
}
\end{figure}

The average sedimentation velocity $V_S$ for different Peclet numbers and system sizes
as a function of hydrodynamic packing fraction $\phi = \frac43 \pi n_c a^3$
is shown in Fig.~\ref{fig1}. The results are normalized by the Stokes velocity
$V_S^0$ (the sedimentation velocity of a single particle in the simulation box),
resulting in the so-called hindered settling function.
At low densities the results are consistent with the result found by Batchelor \cite{Batchelor},
while at higher densities they compare well with a number of other
forms derived for the $\mathrm{Pe}\rightarrow\infty$ limit.
In most experiments the hindered settling function is well described by the
semi-empirical Richardson-Zaki law $V_S/V_S^0 = (1-\phi)^n$, with $n$ ranging between
4.7 and 6.55 \cite{Russ89,Dhont}. Our results fall between these two extremes.
The results compare particularly well with a theoretical prediction by Hayakawa and Ichiki \cite{Haya95},
taking higher order hydrodynamic interactions into account.

One might naively expect that the effect of HI becomes weaker
for $\mathrm{Pe}<1$. Taking into account only Brownian forces would result in
$V_S = V_S^0(1-\phi)$ (because of flux conservation), which heavily underestimates backflow effects.
However, we observe that the results for
\textit{all} Peclet numbers $0.08 \leq \mathrm{Pe} \leq 12$ lie on the
same curve. We emphasise that these results are normalised by the Stokes velocity $V_S^0$ of a single
sphere, which itself decreases with decreasing Peclet number.
The important point is that the \emph{additional} hindrance caused by hydrodynamic interactions is observed to be
unaffected by the actual Pe number. 
A reason for this could be that the \emph{average} sedimentation velocity is determined predominantly by the (time-averaged)
distribution of distances between the colloids. If this is so, then the particle motion generated by the external field must not
lead to a significant change in the microstructure. That this is indeed the case is shown in Fig.~\ref{fig2}, where the main plot shows
the colloidal radial distribution function at volume fraction $\phi = 0.04$ for Peclet numbers 0.2 (stars) and 12 (circles).
For Pe = 0.2 the result is indistinguishable from equilibrium results, and for Pe = 12,
despite the fact that the external field is quite strong, the \emph{average} number of neighbouring particles
at a certain distance from a specific particle changes only very slightly as compared to equilibrium. 
The inset of Fig.~\ref{fig2} shows the structure factor for the same system. 
At Pe = 12, small deviations are found for perpendicular (open circles) and parallel (closed circles) wave vectors,
but again the differences are not very large.
Here we already note that all of these systems are in the unscreened regime.

\section{Spatial correlations in fluctuations}
\label{sec_space}

We next discuss velocity fluctuations around the average.
In colloidal systems the
instantaneous velocity fluctuations $\delta \mathbf{V} =\mathbf{V}
-\mathbf{V}_s$ are
dominated by thermal fluctuations, with a magnitude determined by
equipartition:
\begin{equation}\label{equipartition}
\Delta V_{T}^2 = k_BT/M.
\end{equation}
To disentangle the
hydrodynamic fluctuations from thermal fluctuations, we describe
spatial and temporal correlations in the velocity fluctuations. The
spatial correlation of the $z$ component (parallel to the
sedimentation direction) of the velocity fluctuations can be defined as
\begin{equation}\label{czz}
C_z(\mathbf{r}) \equiv \left \langle \delta V_z(\mathbf{0}) \delta
V_z(\mathbf{r}) \right\rangle\mbox{,}
\end{equation}
where $\left\langle \ldots \right\rangle$ represents an average over time and over
all colloids.
The distance vector $\mathbf{r}$ is taken perpendicular to sedimentation,
$C_z(x)$, or parallel to it, $C_z(z)$.
Note that we will not normalize the correlation functions by their initial values.
Rather, we will normalize them by values which have a more physical meaning, such as
the squared sedimentation velocity $V_S^2$, or the thermal fluctuation
strength $k_BT/M$.

\begin{figure}
  \centerline{\scalebox{0.45}{\includegraphics{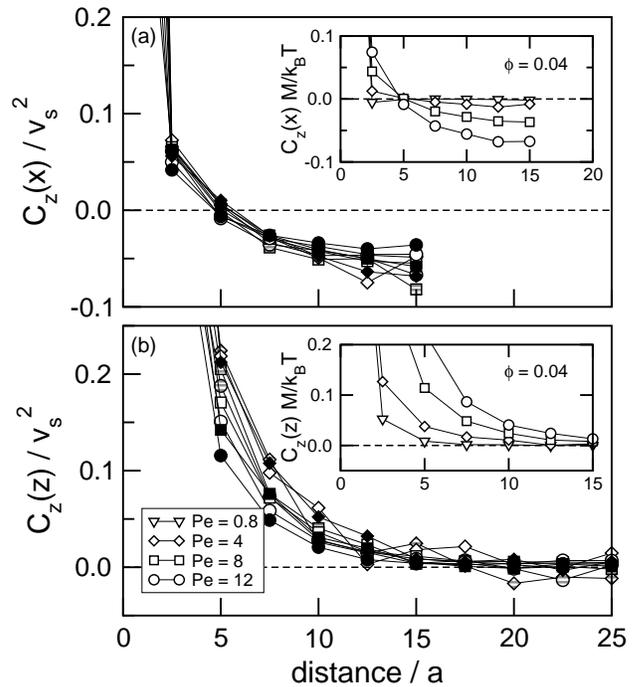}}}
\caption{ 
Spatial correlation functions of the parallel ($z$)
component of the velocity fluctuations as a function of distance
perpendicular (a) and parallel (b) to the external field, for three different
volume fractions ($\phi = 0.02$ (grey symbols), $\phi = 0.04$ (white),
$\phi = 0.086$ (black)) and different Peclet numbers. The correlation functions
are scaled with $V_S^2$ to emphasize hydrodynamic fluctuations.
The insets show how $C_{z}({\bf r})$, scaled with
$k_B T/M$, increases with Pe.
\label{fig3}
}
\end{figure}

\begin{figure}
  \centerline{\scalebox{0.45}{\includegraphics{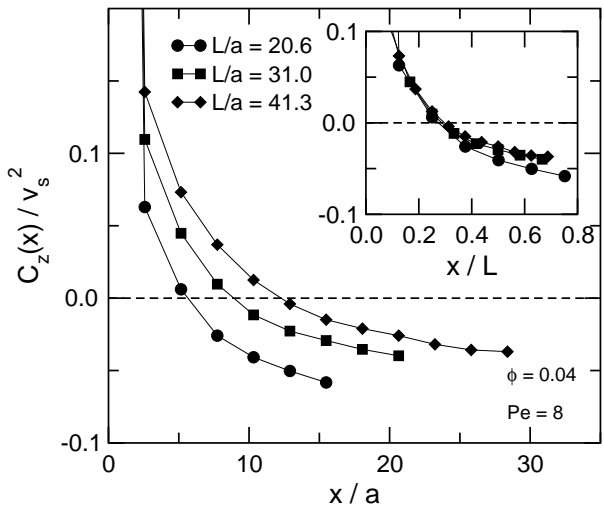}}}
\caption{Spatial correlation functions of the parallel ($z$)
component of the velocity fluctuations as a function of distance
perpendicular to the external field, for different system sizes.
In the inset, distance is scaled with the horizontal box size $L$.
All simulations were performed at $\phi = 0.04$ and $\mathrm{Pe} = 8$.
\label{fig4}
}
\end{figure}

In Fig.~\ref{fig3} we plot $C_z({\bf r})$, which shows a positive
spatial correlation along the direction of flow, and an
anti-correlation perpendicular to the flow, very much like that
observed in the experiments of Nicolai \textit{et al.} \cite{Nico95}. The inset of Fig.~\ref{fig3}(a)
shows that at $\mathrm{Pe} = 0.8$ the correlation in the
perpendicular direction, $C_z(x)$, is almost negligible compared with
the thermal fluctuation strength $k_BT/M$, whereas for larger Pe,
distinct regions of negative amplitude emerge, which grow with
increasing Pe. Similarly, the inset of Fig.~\ref{fig3}(b)
shows correlations in the parallel direction that
rapidly increase with Pe. For the highest Peclet
numbers studied ($4 \leq \mathrm{Pe}
\leq 12$), the amplitudes of these correlations grow
proportionally to $V_S^2$, as shown in the main plots of
Fig.~\ref{fig3}. Unfortunately, because the division by $V_S^2$
amplifies the statistical noise, we are unable to verify whether this
scaling persists for $\mathrm{Pe}< 4$. The
minimum in Fig.~\ref{fig3}(a) is at about half the box width (this
is also the reason why no data points could be collected for $x \geq
15a$). This suggests that the velocity fluctuations are unscreened and
only limited by our box dimensions (see \cite{Ladd1996}). We confirm
this in Fig. \ref{fig4}, where it is seen that 
the correlation length scales linearly with box dimensions.

\section{Temporal correlations in fluctuations}
\label{sec_time}

\begin{figure}
  \centerline{\scalebox{0.45}{\includegraphics{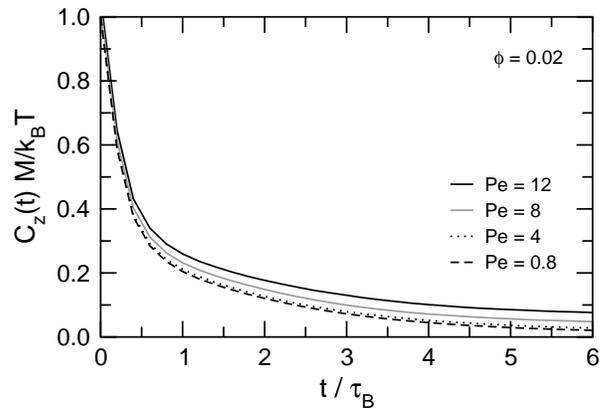}}}
\caption{Temporal correlation functions of the $z$ component
of the velocity fluctuations for $\phi = 0.02$ and different Peclet
numbers. Time is scaled with the Brownian relaxation time
$\tau_B=M/\xi$ and the velocities are scaled with the thermal
fluctuation strength $k_BT/M$.
\label{fig5}}
\end{figure}

\begin{figure}
  \centerline{\scalebox{0.45}{\includegraphics{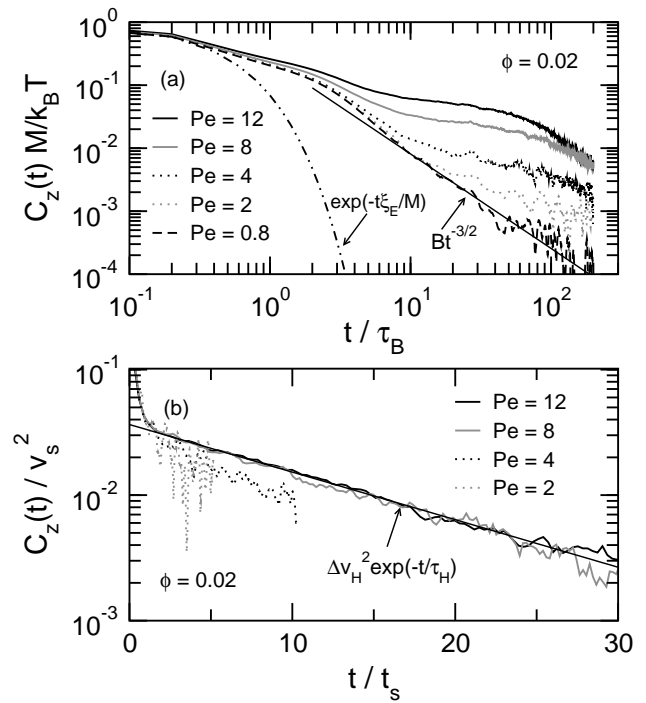}}}
\caption{Temporal correlation functions of the $z$ component
of the velocity fluctuations for $\phi = 0.02$ and different Peclet
numbers.  (a) Time is scaled with the Brownian relaxation time
$\tau_B=M/\xi$ and the velocities are scaled with the thermal
fluctuation strength $k_BT/M$. The straight line is the hydrodynamic
long time tail $B t^{-3/2}$ with  $B^{-1} = 12\rho k_BT(\pi
\nu)^{3/2}$ \protect~\cite{Ernst70}. The
results for $\mathrm{Pe} \leq 1$ are indistinguishable. (b) Time is scaled with
the Stokes time $t_S= a/V_S$ and the velocities are scaled with
$V_S^2$ to highlight hydrodynamic velocity fluctuations.  The straight
line is a fit demonstrating the exponential decay of non-equilibrium
hydrodynamic fluctuations.\label{fig6}}
\end{figure}

Similarly to the spatial correlations of the previous section,
the temporal correlation of the $z$ component of the
velocity fluctuations can be defined as
\begin{equation}\label{czt}
C_z(t) \equiv \left \langle \delta V_z(0) \delta V_z(t) \right\rangle\mbox{,}
\end{equation}
where now $t$ is a correlation time and $\left\langle \ldots \right\rangle$ denotes an average
over all colloids and all time \textit{origins}.
Fig.~\ref{fig5} shows the temporal correlation
functions along the direction of sedimentation on a linear scale. Clearly the correlation is
increasing with increasing Pe number. To investigate this in more detail,
we plot the temporal correlation on a log-log and log-linear plot in Fig.~\ref{fig6}.

At very short times the velocity de-correlation is quantitatively described by Enskog dense-gas
kinetic theory~\cite{Subr75,Hyne77}, which predicts the following decay:
\begin{equation}
\lim_{t \rightarrow 0} C_z(t) = \Delta V_T^2 \exp(-t \xi_E/M), \label{eq_Ens}
\end{equation}
where the Enskog friction coefficient is given by
\begin{equation}
	\xi_E = \frac83 \left( \frac{2\pi k_BT M m_f}{M+m_f} \right)^{1/2} \gamma \sigma_{cf}^2.
\end{equation}
Eq.~(\ref{eq_Ens}) describes the velocity relaxation due to random collisions with the solvent particles.

At intermediate times the temporal correlation follows the well known algebraic long time tail
\begin{equation}
C_{long}(t) = Bt^{-3/2}, \label{eq_long}
\end{equation}
associated with the fact that momentum fluctuations diffuse away at a
finite rate determined by the kinematic viscosity $\nu$.  Analytical
mode-coupling calculations yield a prefactor $B^{-1} = 12 \rho k_BT
(\pi/\nu)^{3/2}$ \cite{Ernst70}. This exactly fits the low Pe ($\leq 1$)
results in Fig.~\ref{fig6}(a) with no adjustable parameters.  We
note that similar agreement was found for the long-time tails for other
parameter choices \cite{Padd05} at equilibrium.  Of course these
simulations are all at finite Pe number, and so are out of
equilibrium, but for small Pe the long-time tail dominates within the
simulation accuracy that we obtain.

\begin{figure}
  \centerline{\scalebox{0.45}{\includegraphics{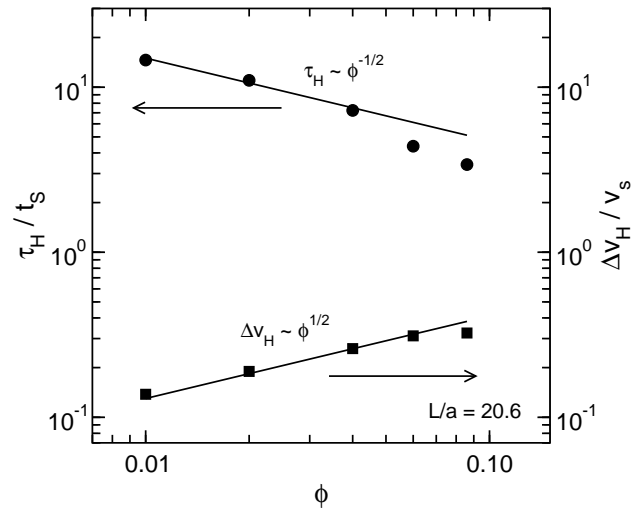}}}
\caption{Scaling of the hydrodynamic relaxation times (left scale) and
 velocity fluctuation amplitudes (right scale) with volume
 fraction. Straight lines are expected scalings for an unscreened
 system\protect~\cite{CaflishLuke,Hinch}.
\label{fig7}
}
\end{figure}

\begin{figure}
  \centerline{\scalebox{0.45}{\includegraphics{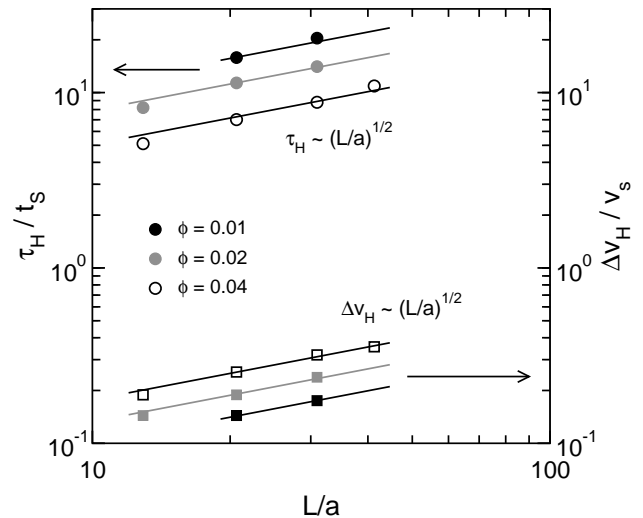}}}
\caption{Scaling of the hydrodynamic relaxation times (left scale) and
 velocity fluctuation amplitudes (right scale) with box size $L$.
 Straight lines are expected scalings for an unscreened
 system\protect~\cite{CaflishLuke,Hinch}.
\label{fig8}
}
\end{figure}

In an experimental study on the sedimentation of non-Brownian
($\mathrm{Pe}
\rightarrow \infty$) particles, Nicolai \textit{et al.} \cite{Nico95} found an exponential temporal relaxation
of the form
\begin{equation}
C_z(t) = \Delta V_H^2 \exp\left( -t/ \tau_H \right).
\label{Chydro}
\end{equation}
This non-equilibrium hydrodynamic effect takes place over much longer
time-scales than the initial exponential relaxation due to random collisions with the solvent particles,
i.e. $\tau_H \gg M/\xi_E$.
The double-logarithmic Fig.~\ref{fig6}(a) shows that a new mode of
fluctuations becomes distinguishable in our simulations for $\mathrm{Pe} >1$.
In the log-linear Fig.~\ref{fig6}(b) the correlation functions are scaled with
$V_S^2$ to highlight the nonequilibrium hydrodynamic fluctuations.
For $\mathrm{Pe} \geq 8$ the fluctuations scale onto a single exponential
master curve, similar to the high-Pe experiments of Nicolai \textit{et al.} \cite{Nico95},
whereas for lower Pe deviations are seen.  From the exponential fit
to Eq.~(\ref{Chydro}), we can estimate the relaxation time $\tau_H$
and the amplitude $\Delta V_H^2$ of the hydrodynamic fluctuations.
These are shown in Fig.~\ref{fig7} for different volume fractions
$\phi$, and in Fig.~\ref{fig8} for different box sizes $L/a$.
The results are consistent with a scaling $\Delta V_H/V_S \propto 
\sqrt{L\phi/a}$ and $\tau_H/t_S \propto \sqrt{L/(\phi a)}$.

These scalings can be understood by a simple heuristic argument by
Hinch {\em et al.} \cite{Hinch}
akin to that used by Caflish and Luke \cite{CaflishLuke}:
Suppose we consider the box volume to consist of two equally large parts, each with a typical linear dimension of $L$.
The average number of colloids in a volume of size $L^3$ is
$\left\langle N \right\rangle = L^3 \phi /
\left( \frac43 \pi a^3 \right)$. Of course the colloids are free to move from one part to the other; the division is
entirely artificial. At low enough volume fraction $\phi$ we assume that the colloidal positions are described
by random Poisson statistics. The typical fluctuation in the number of particles will then be of order
$\sqrt{\left\langle N \right\rangle}$. The extra colloidal weight of order $\sqrt{\left\langle N \right\rangle} M_b g$
in one part of the box causes this part to sediment faster than average.
This is the hydrodynamic fluctuation referred to before. The extra colloidal weight is
balanced by the extra Stokes drag caused by the larger sedimentation velocity, which is of the order of
$6 \pi \eta L \Delta V_H$. Making use of $V_S = M_bg / (6 \pi \eta a)$, we predict for the amplitude of the
hydrodynamic fluctuations:
\begin{equation}\label{eq:VH}
\Delta V_H^2 = V_S^2 \frac{L\phi}{\frac43 \pi a}.
\end{equation}
This is consistent with our observation.
Of course the hydrodynamic fluctuation does not persist indefinitely. It will decorrelate on the order
of the time needed to fall over its own length, for it will then encounter and mix with a region of
average number density. The relaxation time of the hydrodynamic fluctuation is therefore predicted to be
\begin{equation}\label{eq:TH}
\tau_H^2 \approx \frac{L^2}{\Delta V_H^2} = \frac{\frac43 \pi a L^2}
{V_S^2 L \phi} = t_S^2 \frac{\frac43 \pi L}{a \phi},
\end{equation}
where we have used $t_S = a/V_S$.

The above scaling argument does not fix the prefactors. Fitting with the data
in Figs. \ref{fig7} and \ref{fig8} we find
$\Delta V_H \approx 0.29 V_S \left[ \phi (L/a) \right]^{1/2}$ and 
$\tau_H \approx 0.33 t_S \left[ \phi (a/L) \right]^{-1/2}$. It should be noted that
the above results concern the velocity fluctuations parallel to the gravitational
field ($z$). In a similar way we have estimated the perpendicular velocity
fluctuations to be characterized by $\Delta V_{H,\mathrm{perp}} \approx
0.16 V_S \left[ \phi (L/a) \right]^{1/2}$ and  $\tau_{H,\mathrm{perp}} \approx
0.15 t_S \left[ \phi (a/L) \right]^{-1/2}$. Note that the ratio of parallel to 
perpendicular velocity fluctuations is approximately 1.8. This is in agreement with
the experimental low $\phi$ results on non-Brownian spheres by Nicolai \textit{et al.} \cite{Nico95}
and by Segr\'e \textit{et al.} \cite{Segr97},
both of whom observed vertical fluctuations approximately twice the horizontal
fluctuations in the same range of volume fractions.

\section{Diffusion and dispersion}
\label{sec_diffusion}

The equilibrium self-diffusion of a colloidal particle is related to its velocity correlation function
through the following Green-Kubo equation:
\begin{equation}
D_0(t) = \int_0^t \left\langle V_x(\tau) V_x(0) \right\rangle \mathrm{d}\tau, \label{eq_Deq}
\end{equation}
where $V_x$ is a Cartesian component of the colloidal velocity. For
large enough times $t$ the integral $D_0(t)$ converges to the
equilibrium self-diffusion coefficient $D_0$.

\begin{figure}
  \centerline{\scalebox{0.45}{\includegraphics{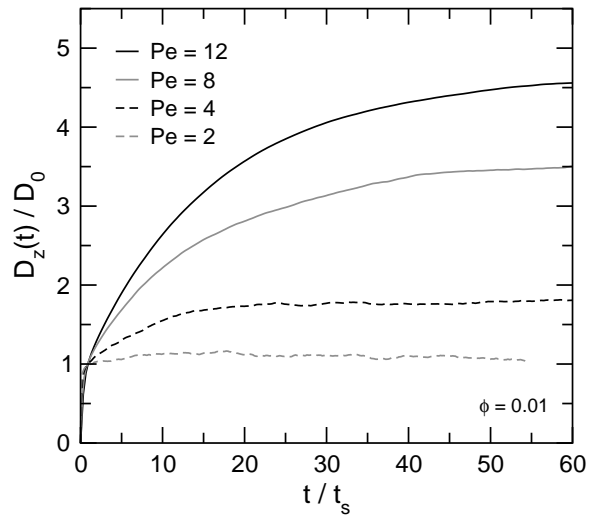}}}
\caption{Time dependent self-diffusion coefficient parallel to gravity for different
Pe numbers. $D_0$ is the equilibrium self-diffusion coefficient. \label{fig9}
}
\end{figure}

During sedimentation, the diffusion is enhanced by the hydrodynamic
fluctuations. In fact, the diffusion is no longer isotropic but
tensorial. Focusing first on the component parallel to gravity, we
define the parallel diffusion coefficient similarly to
Eq.~(\ref{eq_Deq}) as the large time limit of
\begin{equation}
D_z(t) = \int_0^t \left\langle \delta V_z(\tau) \delta V_z(0) \right\rangle \mathrm{d}\tau.
\end{equation}
In Fig.~\ref{fig9} we show $D_z(t)$ normalized by the
equilibrium value $D_0$ for a range of Pe numbers.  Note that even
though the hydrodynamic fluctuations may be small compared to $C(0)$,
they nevertheless have a significant contribution to the diffusivity
because the time-scale $\tau_H$ is much longer than $\tau_\nu$.

\begin{figure}
  \centerline{\scalebox{0.45}{\includegraphics{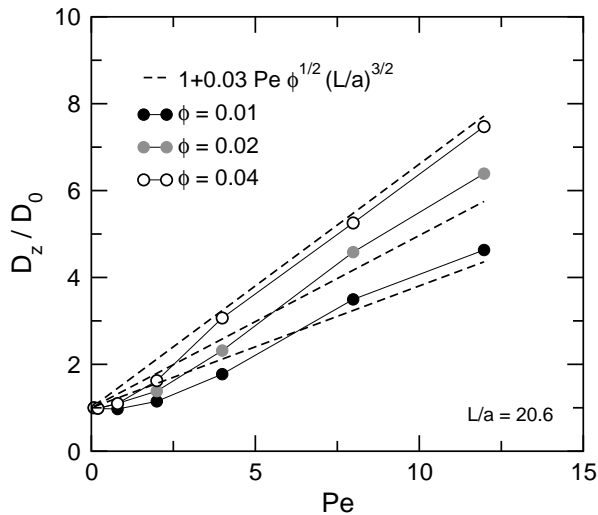}}}
\caption{Self-diffusion coefficient parallel to gravity versus Peclet number for different
concentrations $\phi$. $D_0$ is the equilibrium self-diffusion coefficient. Dashed lines
are predictions from Eq.~(\ref{eq_diffusion_Pe}).
\label{fig10}
}
\end{figure}

To understand the total diffusivity, we make the following addition approximation:
\begin{equation}
D_z = D_0 + D_H
\end{equation}
where $D_0$ is equilibrium diffusion coefficient and $D_H$ the dispersion due to
non-equilibrium hydrodynamic fluctuations. The former can be approximated as a 
sum of Stokes and Enskog diffusion coefficients, see \cite{Padd06}.
The non-equilibrium hydrodynamic dispersion can be estimated using the previous
scaling arguments:
\begin{equation}
D_H  \approx  \Delta V_H^2 \tau_H 
    \propto  V_S a \phi^\frac12 \left( \frac{L}{a} \right)^{3/2}
\end{equation}
Taking the prefactors found in the previous section, and
rewriting $V_S a$ as $\mathrm{Pe} D_0$, we therefore predict
\begin{equation}
D_z = D_0 \left[ 1 + 0.03\ \mathrm{Pe}\ \phi^{1/2} \left(\frac{L}{a}\right)^{3/2} \right]
\label{eq_diffusion_Pe}
\end{equation}
for small enough, i.e. unscreened, systems. For small Pe ($<1$) the self-diffusion
coefficient is largely independent of Pe and equal to $D_0$,
whereas for very large Pe ($\gg 1$) it becomes proportional to Pe. 
This is confirmed in Fig. \ref{fig10} where the dashed lines show the Pe and $\phi$ dependence
of Eq.~\ref{eq_diffusion_Pe}.

The diffusion in the plane perpendicular to gravity is also enhanced by the
hydrodynamic fluctuations, similar to Eq.~(\ref{eq_diffusion_Pe}), but with a smaller
prefactor of 0.004 instead of 0.03 (not shown). The ratio of hydrodynamic diffusivities,
$D_H / D_{H,\mathrm{perp}} \approx 7$ is similar to what is found in the experiments
of Nicolai \textit{et al.} \cite{Nico95} for non-Brownian spheres.

Although our simulations are in the unscreened regime, it is
interesting to also consider the hydrodynamic contribution to the
diffusion coefficient in the screened regime.  If we apply the
experimental fits of Segr\'e \textit{et al.}\
\cite{Segr97} for $\Delta V_H$ and the correlation length $\xi$, then
the simple scaling arguments above suggest that
\begin{equation}
D_H \propto \mathrm{Pe} \, D_0,
\end{equation}
which is {\it independent} of $\phi$.  The exact pre-factor is hard to
determine in the screened regime. Nevertheless, an estimate can be
made if we assume that $\tau_H$ has the same pre-factor in the
experiments as we find in our simulations. For example, if we replace
$L/2$,which measures the location of the minimum of the perpendicular
correlation functions, with $\xi_\mathrm{perp}$, its value for the
screened regime\cite{Segr97}, then we find $D_{H,\mathrm{perp}}/D_0
\approx 1.1$ Pe.  For $D_{H,\mathrm{parallel}}/D_0$ we expect a
pre-factor several times larger.  In the screened regime the
hydrodynamic contributions to the diffusion should dominate for Pe$
\gtrsim 1$.  In practice however, we expect that for many colloidal
dispersions effects such as polydispersity\cite{Nguy05} may temper the
size of the swirls, and thus reduce the hydrodynamic contribution to
diffusion.

\section{Finite sedimentation in a horizontal planar slit}
\label{sec_finite}

Up to this point we have focused on steady-state sedimentation by
applying periodic boundary conditions and giving the system enough
time to overcome transient flow effects.

One may wonder what happens if the particles are confined and are not
given enough time to reach steady-state.  Very recently, Royall
\textit{et al.} \cite{Royall07} studied nonequilibrium sedimentation
of colloids in a horizontal planar slit, at a Peclet number of order 1,
using laser scanning confocal microscopy. Among other things, they
measured the time evolution of the one-dimensional colloid density
profile $\rho(z,t)$, where the $z$-axis is normal to the horizontal
plane.  Two cases were considered. In the first case an initially
homogenised sample was allowed to sediment to the bottom of the
capillary. Good agreement was found with a dynamical density functional
theory (DDFT) calculation that included a density-dependent mobility function. In
the second case they considered an equilibrated sample turned upside
down so that the previous sediment suddenly finds itself
at the top of the capillary. In this case sedimentation
proceeds in an entirely different fashion. A strong finger-like
inhomogeneity was observed, accompanied by maze-like lateral pattern
formation.

\begin{figure}
  \centerline{\scalebox{1.0}{\includegraphics{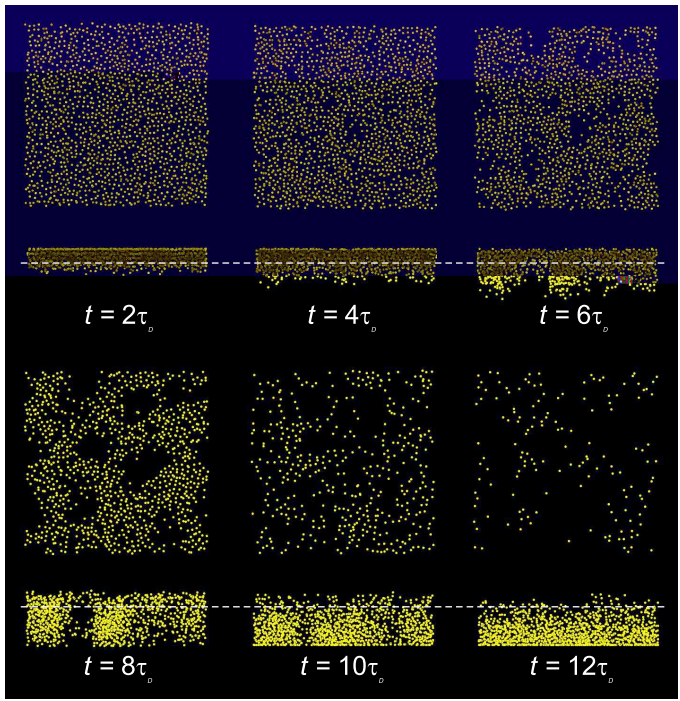}}}
\caption{An equilibrated sediment in a planar slit is turned upside down and allowed to sediment at Pe = 4. Shown here are
the horizontal $xy$-plane and the corresponding vertical $xz$-plane at 6 different times after field reversal. 
The dashed line indicates the height $z$ where the snapshots of the $xy$-plane are taken.
A strong fingerlike inhomogeneity develops quickly, accompanied by maze-like lateral pattern formation.
\label{fig11}
}
\end{figure}

Inspired by these experiments, we set up a box of size $180 \times 180
\times 60 a_0$ ($116 \times 116 \times 39 a$), with periodic
boundaries in the $x$ and $y$ direction, and with walls at the top and
bottom in the $z$ direction. (This corresponds to a height of about
$32 a $, close to the experimental value of $36 a$.)  We add 6500
colloids $(\phi \approx 0.06$) and apply an external field upwards
such that Pe = 4. After reaching the equilibrium distribution, at $t =
0$ we suddenly reverse the field, again at Pe = 4. We observe a
maze-like lateral pattern, Fig.~\ref{fig11}, which shows striking
similarities to the experimental observations~\cite{Royall07}.
The characteristic length of the maze-like lateral pattern is approximately
equal to the height of the slit.
%
%
 It has been suggested \cite{Royall07} that there may be a relation
 between this phenomenon and the swirls observed in steady-state
 sedimentation, but also that the swirls are reminiscent of a
 Rayleigh-Taylor type instability in two layered liquids, with the
 steep initial density gradient resembling a (very diffuse)
 liquid-liquid interface.  With the current data, we cannot
 conclusively determine the origin of this instability.  Nevertheless
 it is gratifying that our simulations produce such similar, and
 nontrivial, behaviour as the experiments under similar conditions.
This can be viewed as an additional validation of our simulation model.

\begin{figure}
  \centerline{\scalebox{0.5}{\includegraphics{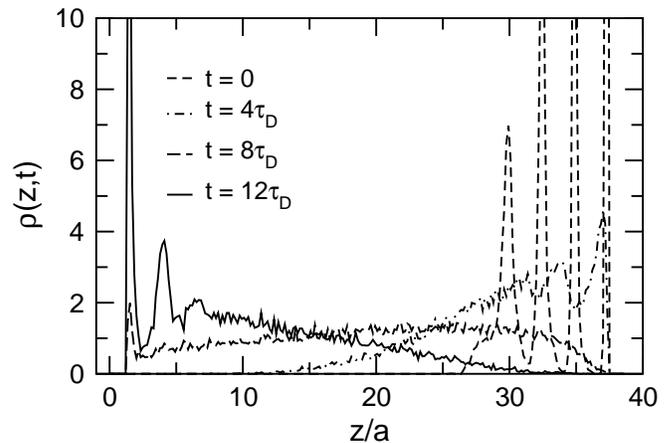}}}
\caption{Time evolution (right to left) of the one-dimensional density profile for sedimentation in a horizontal slit.
$\rho$ is normalised such that it equals 1 for a homogeneously filled
slit.  The final state (on the left) closely resembles the initial
state (on the right), but is not shown for clarity.
\label{fig12}
}
\end{figure}

In Fig.~\ref{fig12} we analyse the time evolution of the
one-dimensional density profile $\rho(z,t)$.  The crystal-like layers
at the top plate for $t < 0$, disappear and then reappear again at the
bottom of the plate.  It would be interesting to compare these results
to calculations using DDFT.  Since the latter technique does not explicitly contain any
long-ranged hydrodynamics, one would expect it to have difficulty in
reproducing 
the swirls observed in the simulations and experiments.  Nevertheless,
because both the initial and final states are constrained by
equilibrium statistical mechanics (for which DFT is very accurate),
the one-body density $\rho(z,t)$ may not be a very sensitive measure
of the more complex dynamics that arise from hydrodynamics.


\section{Discussion}
\label{sec_discussion}

As seen in Fig.~\ref{fig6}, the short time velocity fluctuations are
dominated by thermal fluctuations at all Peclet numbers studied.
The relative strength of the $t=0$ thermal and hydrodynamic velocity 
fluctuations follows from simple scaling relations.  
Using 
\begin{equation}
\mathrm{Re \, Pe} = \frac{(V_S a)^2}{D_0 \nu},
\end{equation}
which follows from the definitions of Pe and Re, together with
Eqs.~(\ref{equipartition}) and~(\ref{eq:VH}),
the following  relationship between hydrodynamic and thermal fluctuations emerges:
\begin{equation}
\frac{\Delta V_H^2}{\Delta V_T^2} \approx  \alpha\, \mathrm{Re\, Pe}\, \phi \frac{L}{a}  \frac{\rho_c}{\rho} \qquad \mbox{(unscreened)}, 
\end{equation}
where the simplifying assumption that $M_c \approx \frac43 \pi \rho_c a^3$,
with $a$ the hydrodynamic rather than the physical radius, was also
made.  The numerical pre-factor $\alpha$ is small and can be extracted
from Fig.~\ref{fig8} to be $\alpha \approx 0.05$ for fluctuations
parallel to the flow, and $\alpha \approx 0.015$ for fluctuations
perpendicular to the flow.

The above scaling holds for the unscreened regime; in the screened
regime the ratio of $V_H$ to $V_T$ will be smaller. Consider, for example,
 the experimental results of Segr\'e \textit{et al.} \cite{Segr97}. If we
  take their fits to the scaling of the parallel fluctuations in the
  screened regime, together with the estimates $\mathrm{Re} = 5\times
  10^{-5}$ and $\mathrm{Pe} \approx 2000$, the scaling becomes:
\begin{equation}
\frac{\Delta V_H^2}{\Delta V_T^2} \approx  \phi^{2/3} \mathrm{Re\, Pe} \qquad \mbox{(screened)}
\end{equation}
 for flows in the parallel direction. This suggests that this ratio is
 small in the experiments, from $2 \times 10^{-4}$ for $\phi=10^{-4}$
 to $0.02$ for $\phi=0.1$. So despite the fact that the Pe number in
 these experiments appears to be high, there is no need for an
 effective gravitational ``temperature'' \cite{Segr01} to thermalise:
 at short correlation times the usual thermal fluctuations are still
 dominant. However, because the product $\mathrm{Re\, Pe}$ scales with
 quite a high power of $a$, as fast as $a^7$, the ratio $\Delta
 V_H^2/\Delta V_T^2$ will increase rapidly for larger particles and
 gravitational temperature will become essential for thermalisation.

When comparing parallel and perpendicular components it is important
to mention that in numerical works where thermal fluctuations are
neglected very strong anisotropies in velocity fluctuations,
hydrodynamic relaxation times, and diffusivities are often found. For
example Ladd \cite{Ladd97} finds $D_H / D_{H,\mathrm{perp}} \approx 25$ in his
Lattice Boltzmann simulations. This was attributed to
periodic boundary conditions. However, we also use periodic boundary
conditions and find results much closer to experimental results (a
diffusivity ratio of $\sim 7$). We therefore conclude that thermal
fluctuations reduce the anisotropy. This could be tested in Lattice
Boltzmann simulations by adding fluctuating stress \cite{Cates}.

\section{Conclusion}
\label{sec_conclusion}

In conclusion, we have studied the interplay of hydrodynamic and
thermal fluctuations using a novel simulation technique. The two types
of fluctuations appear to act independently, at least in the
unscreened regime. We find that hydrodynamic interactions are
important for the average sedimentation velocity for Peclet numbers as
low as 0.08, whereas thermal fluctuations may remain important up to
very large Peclet numbers. Neither may be ignored for a significant
range of Peclet numbers. We also calculate the hydrodynamic contributions
to the diffusion coefficient, and find that with increasing Pe number they
rapidly become much larger than the equilibrium diffusion
coefficient. As an additional test of the method we studied finite sedimentation in a horizontal slit,
and found characteristic lateral patterns in agreement with recent experiments.

\acknowledgments
J.T.P. thanks the Netherlands Organisation for Scientific
Research (NWO), and A.A.L. thanks the Royal Society (London) for financial support.
We thank J. Hinch, R. Bruinsma, S. Ramaswamy, I. Pagonabarraga, W. Briels, and C.P. Royall
for very helpful conversations.


\end{document}